# Quantum Hall Effect on Top and Bottom Surface States of Topological Insulator $(Bi_{1-x}Sb_x)_2Te_3$ Films

R. Yoshimi[1*], A. Tsukazaki[2,3], Y. Kozuka[1], J. Falson[1], K. S. Takahashi[4], J. G. Checkelsky[1],

N. Nagaosa[1,4], M. Kawasaki[1,4] and Y. Tokura[1,4]

[1] *Department of Applied Physics and Quantum-Phase Electronics Center (QPEC),*

*University of Tokyo, Tokyo 113-8656, Japan*

[2] *Institute for Materials Research, Tohoku University, Sendai 980-8577, Japan*

[3] *PRESTO, Japan Science and Technology Agency (JST), Chiyoda-ku, Tokyo 102-0075,*

*Japan*

[4] *RIKEN Center for Emergent Matter Science (CEMS), Wako 351-0198, Japan.*

* Corresponding author: yoshimi@cmr.t.u-tokyo.ac.jp

**The three-dimensional (3D) topological insulator (TI) is a novel state of matter as characterized by two-dimensional (2D) metallic Dirac states on its surface[1-4]. Bi-based chalcogenides such as $Bi_2Se_3$, $Bi_2Te_3$, $Sb_2Te_3$ and their combined/mixed compounds like $Bi_2Se_2Te$ and $(Bi_{1-x}Sb_x)_2Te_3$ are typical members of 3D-TIs which have been intensively studied in forms of bulk single crystals and thin films to verify the topological nature of the surface states[5-12]. Here, we report the realization of the Quantum Hall effect (QHE) on the surface Dirac states in $(Bi_{1-x}Sb_x)_2Te_3$ films ($x = 0.84$ and 0.88). With electrostatic gate-tuning of the Fermi level in the bulk band gap under magnetic fields, the quantum Hall states with filling factor $\nu = \pm 1$ are resolved with quantized Hall resistance of $R_{yx} = \pm h/e^2$ and zero longitudinal resistance, owing to chiral edge modes at top/bottom surface Dirac states. Furthermore, the appearance of a $\nu = 0$ state ($\sigma_{xy} = 0$) reflects a pseudo-spin Hall insulator state when the Fermi level is tuned in between the energy levels of the non-degenerate top and bottom surface Dirac points. The observation of the QHE in 3D TI films may pave a way toward TI-based electronics.**

Quantum transport in Dirac electron systems has been attracting much attention for the half-integer quantum Hall effect (QHE), as typically observed in graphene[13, 14]. The recently discovered topological insulator (TI) possesses metallic Dirac states on the edge or surface of an insulating bulk[1-4]. With the application of a magnetic field ($B$) the unique features of Dirac bands may be exemplified via the formation of Landau levels (LLs). The QHE is the hallmark of dissipationless topological quantum transport originating from one-dimensional chiral edge modes driven by cyclotron motion of two-dimensional (2D) electrons. Unlike the case of graphene which has a four-fold degeneracy originating from the spin and valley degrees of freedom, the degeneracy is completely lifted in the spin-polarized Dirac state of 2D and 3D TIs. For such 3D TI films the top and bottom surfaces support surface states with opposite spin-momentum locked modes when top and bottom surfaces are regarded as two independent systems. Such a helicity degree of freedom in real space can be viewed as the pseudo-spin variable and is hence expected to yield a new quantum state via tuning of surface magnetism and/or Fermi level ($E_F$) that is applicable to quantum computation functions[15-17]. Although intensive research has been carried out for bulk crystals, thin films, and field-effect devices[5-12], parasitic bulk conduction and/or disorder in the devices continues to hamper efforts to resolve quantum transport characteristics of the Dirac states on chalcogenide 3D TIs surfaces. The most venerable example of the QHE with least bulk conduction has been achieved in 70nm strained HgTe film[18]. Compared to the HgTe system, 3D-TIs of Bi-chalcogenides such as $Bi_2Se_2Te$ and $(Bi_{1-x}Sb_x)_2Te_3$ have a good potential for exploring the Dirac surface states with wide controllability of transport parameters (resistivity, carrier type and density) and band parameters (energy gap, position of Dirac point and Fermi velocity) by changing the compositions[19, 20].

In this work, we report transport measurements on 3D TI thin films of $(Bi_{1-x}Sb_x)_2Te_3$ ($x$ = 0.84 and 0.88; both 8 nm thick). Using molecular beam epitaxy[19] and insulating InP (111) substrates, the $E_F$ of as-grown film was tuned near to the bulk band edge by precisely controlling the Bi/Sb composition ratio in the film[19, 20]. Films were then fabricated into photolithography defined gated Hall-bar devices to allow electrostatic tuning of $E_F$. A cross-sectional schematic of the device structure and the top-view image are shown in Figs. 1a and 1b, respectively. The device consists of Hall bar defined by Ar ion-milling, and an atomic-layer-deposited $AlO_x$ insulator isolates Ti/Au top gate with electron-beam evaporated Ti/Au electrodes (see Methods). The magnetotransport measurements were carried out in a dilution refrigerator by low-frequency (3 Hz) lock-in technique with low excitation current of 1 nA to suppress heating.

First, device operation was examined at $B = 0$ T. Figure 1c shows the electric field effect controlled conductivity $\sigma_{xx}$ of the $x = 0.84$ film as a function of top gate voltage $V_G$ with changing temperature. The minimum conductance of roughly 2 $e^2/h$ ($e$: the elemental charge, $h$: Planck constant) is observed at $V_G = -1.7$ V. On both sides of this minimum, $\sigma_{xx}$ shows a linear but asymmetric increase with increasing or decreasing $V_G$. The weak temperature dependence of $\sigma_{xx}$ for a wide range of $V_G$ is characteristic of the gapless nature of Dirac states under finite disorder[21]. Thus we ascribe the conductivity of this TI film below 1 K to the Dirac surface states with a small contribution from bulk conduction. To verify the ambipolar nature of the device, the longitudinal and transverse resistance $R_{xx}$ and $R_{yx}$ as a function of $V_G$ at $B = 3$ T were measured (Fig. 1d). As expected, this results in a sign change of $R_{yx}$ at a certain $V_G$ which we hereafter define as the gate voltage corresponding to $E_F$ being located at the charge neutral point (CNP), $V_{CNP}$. This point coincides closely with the $V_G$ at which $R_{xx}$ reaches a maximum. At the $V_{CNP}$, the Hall effects from top and bottom surface states appear to cancel, resulting in $R_{yx} \sim 0$, although electron-rich and hole-rich puddles are thought to still exist, as in the case of graphene[22]. To capture the essence of the observed phenomena, we hereafter take the working hypothesis that $E_F$ shifts equally on top and bottom states which in turn retain the difference in energy position. The inverse of Hall coefficient $1/R_H$ is shown in Fig. 1e, which would be proportional to 2D charge carrier density $n_{2D} = 1/eR_H$, in the simplest case. Asymmetric behavior of $1/R_H$ between positive and negative $V_G$ regions with respect to $V_{CNP}$ is in accord with the asymmetric $\sigma_{xx}$ behavior at zero magnetic field as shown in Fig. 1c. The more efficient increase in $\sigma_{xx}$ with electron accumulation (positive $V_G$) is often observed in ambipolar TI transistors[8-12] and may be related to the difference in $v_F$ (Fermi velocity) above and below the Dirac point[19]. In addition, the proximity of the Dirac point to the valence band edge needs to be taken into account for the present thin film system; in most of the negative region of $V_G - V_{CNP}$, $E_F$ must go inside the valence band in which doped holes appears to be fully localized at low temperatures below 1 K, as argued later.

With applying higher magnetic field of $B = 14$ T, clear signatures of QHE are revealed in the temperature dependence of $R_{xx}$ and $R_{yx}$, as shown in Figs. 2a-d for the $x = 0.84$ film and in Figs. 2e-h for the $x = 0.88$ film. We first focus on the results of the $x = 0.84$ film. $R_{xx}$ increases steeply with lowering the temperature for $V_G$ corresponding to the CNP, while decreases rapidly toward zero in the both sides at higher and lower $V_G$. Concomitantly to the decrease in $R_{xx}$, $R_{yx}$ reaches the values of quantum resistance $\pm h/e^2 = \pm 25.8$ kΩ and forms plateaus in $V_G$ at $T = 40$ mK. This correspondence between the rapidly declining $R_{xx}$ value and the quantized $R_{yx}$ plateaus at $\pm h/e^2$ are distinct evidence for QHE at Landau filling factor $\nu = \pm 1$, respectively, as schematically shown in Fig. 2i. The LL splitting energy of Dirac dispersion ($E_n$) is given by $E_n = \text{sgn}(n)\sqrt{2|n|\hbar v_F^2 eB}$, where $n$ is the LL index.

Using the data of $R_{xx}$ and $R_{yx}$, $\sigma_{xy}$ and $\sigma_{xx}$ as functions of $V_G - V_{CNP}$ for the $x = 0.84$ film are plotted in Figs. 2c and 2d. Again, the Hall plateaus at $\sigma_{xy} = \pm e^2/h$ as well as minima of $\sigma_{xx}$ approaching zero (black triangles) are observed and are indicative of the QHE with $\nu = \pm 1$. In this plot, however, two additional features are to be noted. The first is an unexpectedly wide $\sigma_{xy}$ plateau and thermally activated behavior of $\sigma_{xx}$ for the $\nu = +1$ ($\sigma_{xy} = e^2/h$) state in the corresponding $V_G - V_{CNP}$ (negative) region. As already noted in the $V_G$ asymmetric change of $\sigma_{xx}$ (Fig. 1c) in the negative region of $V_G - V_{CNP}$, i.e. hole-doping, $E_F$ readily reaches the valence band top. The energy position of Dirac point of the $x = 0.84$ film lies by at most 30 meV above the valence band top, while the LL splitting between $n = 0$ and $n = -1$ levels amounts to 70 meV at 14 T, according to resonant tunneling spectroscopy on similarly grown thin films of $(Bi_{1-x}Sb_x)_2Te_3$ (Ref. 19). From the consideration of the Fermi velocity of Dirac cone, the $V_G - V_{CNP}$ value at which the $E_F$ reaches the valence band top is estimated to be around −1.5 V or an even a smaller absolute value. Therefore, in the $V_G - V_{CNP}$ region where the quantum Hall plateau or its precursor is observed, the $E_F$ locating between $n = 0$ and $n = -1$ LLs of the surface state is close to or already buried in the valence band, as schematically shown in Fig. 2i. While the doped but localized holes in the valence band may hardly contribute to transport, *i.e.* $\sigma_{xy}$ (bulk) $\ll \sigma_{xy}$ (surface), the relative $E_F$ shift with negatively sweeping $V_G - V_{CNP}$ becomes much slower as compared with the positive sweep case owing to the dominant density of states of the valence band. This explains a wider plateau region for $\nu = +1$ in the hole-doping side, contrary to the normal behavior of electron accumulation side, $\nu = -1$.

The second notable feature in Fig. 2c is the emergence of the $\nu = 0$ state around $V_G = V_{CNP}$, as seen in the step of $\sigma_{xy}$ and (finite) minimum in $\sigma_{xx}$ as functions of $V_G - V_{CNP}$. This state is more clearly resolved in the $x = 0.88$ film, as shown in Figs. 2e-h, on which we focus hereafter. In a similar manner to the $x = 0.84$ film, the $x = 0.88$ film also shows with lowering temperature divergent behavior of $R_{xx}$ around $V_G = V_{CNP}$, while approaching zero around $V_G - V_{CNP} = -1.5$ V. The $R_{yx}$ reaches 25.8 k$\Omega$ around $V_G - V_{CNP} = -1.5$ V forming the $\nu = +1$ quantum Hall state, while in the electron doping regime $R_{yx}$ reaches $-20$ k$\Omega$, short of the quantized value. The failure to form the fully quantized $\nu = -1$ state is perhaps related to the disorder of the surface Dirac state which is induced by the compositional/structural disorder of the as-grown film and cannot be overcome by gate tuning.

Nevertheless, the $\nu = 0$ feature is clearly resolved for the $x = 0.88$ film, as shown in the $V_G$ dependence of $\sigma_{xy}$ (Fig. 2g) calculated from $R_{xx}$ and $R_{yx}$. In addition to the $\sigma_{xy} = e^2/h$ ($\nu = +1$) plateau, a plateau at $\sigma_{xy} = 0$ appears at around $V_G = V_{CNP}$. The plateau broadening occurs via centering at $\sigma_{xy} = 0.5\ e^2/h$ as the isosbestic point with elevating temperature. In accordance with the plateaus in $\sigma_{xy}$, $\sigma_{xx}$ takes a minima at $\nu = +1$ and 0, as shown in Fig. 2h. Here, we can consider the contribution of the both top and bottom surface Dirac states to this quantization of $\sigma_{xy}$, as schematically shown in Fig. 2j. At the $\nu = +1$ ($\nu = -1$) state, the both top and bottom surfaces are accumulated by holes (electrons) with $E_F$ being located between $n = 0$ and $n = -1$ ($n = +1$) LLs, giving rise to the chiral edge channel. In contrast, we assign the $\nu = 0$ state to the gapping of the chiral edge channel as the cancelation of the contributions to $\sigma_{xy}$ from the top and bottom surface states with $\nu = \pm 1/2$, when $E_F$ locates in between the energy levels of the top and bottom surface Dirac points ($n = 0$ levels), as shown in Fig. 2j. This $\nu = 0$ state can hence be viewed as a *pseudo-spin Hall insulator*, if we consider the top and bottom degree of freedom as the pseudo-spin variable. Such an observation of a zero conductance plateau has been reported also in disordered graphene under very high magnetic field[23-26] and analyzed theoretically[27], as well as in the 2D TIs, the quantum wells of HgTe[28] and InAs/GaSb[29]. From the analyses shown in the following, we propose here that the major origin for the presence of $\sigma_{xy} = 0$ is more like the energy difference of the top/bottom Dirac points rather than other effect such as electron-hole puddles in composition inhomogeneity.

To further discuss the characteristics of these QH states, we investigate the $B$ dependence of $\sigma_{xy}$ (Figs. 3a and b). The analysis of the plateau width against $V_G$ determines the phase diagram as shown in Figs. 3c and d. The plateau edges are determined from the second derivative of $\sigma_{xy}$ with respective to $V_G$ (see Supplementary Information (SI)), while the plateau transition points between $\nu = 0$ and $\nu = \pm 1$ are defined here by $\sigma_{xy} = \pm 1/2$ $e^2/h$. The plateau shrinks with decreasing $B$ for the $\nu = -1$ state of the $x = 0.84$ film (Fig. 3a). However, the $\nu = +1$ state for the both films appears to be rather robust with reducing $V_G$ (doping more holes), since $E_F$ positions already below the top of the valence band, perhaps for $V_G - V_{CNP} < -1.5$ V. On the other hand, the $\nu = 0$ plateau is only weakly dependent on $B$, although the plateau width is wider for $x = 0.88$ than for $x = 0.84$. The observation of $\nu = 0$ requires the condition that the Fermi level is located in between the energy levels of the top and bottom surface Dirac points (Fig. 2j). From the Hall data in the relatively high positive $V_G - V_{CNP}$ region (electron-doping) shown in Fig. 1e, we can know the relation between the sum of the top and bottom Dirac electron density versus $V_G - V_{CNP}$. Then, with the values of the $\nu = 0$ plateau width between the $\sigma_{xy} = \pm 0.5\ e^2/h$ points ($\delta V_G$ ~0.9 V and 1.4 V; see Figs. 3c and d) and the Fermi velocity ($v_F \sim 5 \times 10^5$ m/s)[19], we can estimate the energy difference ($\delta E_{DP}$) between the Dirac points at the top and bottom surface states to be ~ 50 meV and ~70 meV for the $x = 0.84$ and $x = 0.88$ film, respectively (see SI). These values should be compared with a much larger band gap energy (~ 250 meV). The energy difference $\delta E_{DP}$ is, however, considerably larger than a Zeeman shift (~9 meV at 14 T)[19], which rationalizes the above analysis with ignoring the Zeeman shift of the $n = 0$ LL. While the reason why the two films ($x = 0.84$ and $0.88$) show such a difference in $\delta E_{DP}$ is not clear at the moment, we speculate that the monolayer buffer layer of $Sb_2Te_3$ ($x = 1.0$) used for the growth of the $x = 0.88$ film (see SI) may cause the considerably higher energy position of the Dirac point at the bottom surface. Incidentally, for the region of $|V_G - V_{CNP}| < 1/2\ \delta V_G$, electron accumulation at the top surface and hole accumulation at the bottom surface should coexist. This may naturally explain the observed (Fig. 1e and see also SI) deviation from the linear relationship between $1/R_H$ and $V_G - V_{CNP}$ as well as the extrema of $1/R_H$ observed at around $\pm 1/2 \delta V_G$.

Figure 4 summarizes the flow of conductivity tensor ($\sigma_{xy}$, $\sigma_{xx}$) plotted with the two experimental subparameters ($T$ and $V_G$) at 14 T. With decreasing $T$, the flow in ($\sigma_{xy}$, $\sigma_{xx}$) tends to converge toward either of ($\sigma_{xy}$, $\sigma_{xx}$) = ($-e^2/h$, 0), (0, 0) or ($e^2/h$, 0) at high magnetic field (e.g. 14 T), which corresponds to $\nu = -1$, 0 and +1 QH state, respectively. Incipient convergence to $\nu = 0$ is discerned for the $x = 0.84$ film, while the $\nu = -1$ state is not discernible for the $x = 0.88$ film. Among these three QH states, the unstable fixed point appears to lie on the line of $\sigma_{xy} = \pm 0.5\, e^2/h$ (approximately with the critical $\sigma_{xx}$ value of ~ 0.5 $e^2/h$) which corresponds to the crossing of $E_F$ at the $n = 0$ LL (or Dirac point) of the bottom and top surface state (see Fig. 2j), respectively[30].

In conclusion, we have successfully observed the QHE at $\nu = \pm 1$ and 0 states in three-dimensional TI thin films of $(Bi_{1-x}Sb_x)_2Te_3$ ($x = 0.84$ and 0.88). Due to a considerable difference of the Dirac point (or $n = 0$ LL) energies of the top and bottom surfaces of the thin film, the $\nu = 0$ state observed at $\sigma_{xy} = 0$ is interpreted as a pseudo-spin Hall insulator with the top/bottom degree of freedom as the pseudo-spin. Further studies on non-local transport in mesoscopic structures will open the door to dissipationless topological-edge electronics based on the 3D topological insulators.

**Methods**

Thin films of $(Bi_{1-x}Sb_x)_2Te_3$ ($x$ = 0.84 and 0.88) were fabricated by molecular beam epitaxy (MBE) on semi-insulating InP (111) substrate. The Bi/Sb composition ratio was calibrated by the beam equivalent pressure of Bi and Sb, namely $8 \times 10^{-7}$ Pa and $4.2 \times 10^{-6}$ Pa for $x$ = 0.84 and $6 \times 10^{-7}$ Pa and $4.4 \times 10^{-6}$ Pa for $x$ = 0.88. The Te flux was over-supplied with the Te / (Bi + Sb) ratio kept at about 20. The substrate temperature was 200°C and the growth rate was about 0.2 nm/min. Fabrication procedures for the $x$ = 0.84 and 0.88 films are slightly different at the initial growth on InP surfaces. We grew the 0.84 film with supplying Te and (Bi + Sb) from the initial stage. For the $x$ = 0.88 film, we started with supplying Te and Sb for a monolayer growth of $Sb_2Te_3$ buffer layer followed by Bi shutter opening. This difference may be an origin of the larger energy gap $\delta E_{DP}$ of the Dirac points between the top and bottom surfaces (Fig. 2j) for the $x$=0.88 film. After the epitaxial growth of 8nm-thick thin films, those were annealed *in-situ* at 380°C to make the surface smoother. $AlO_x$ capping layer was deposited at room temperature with an atomic layer deposition system immediately after the discharge of the samples from MBE. This process turned out to be effective to protect the surface from the degradation. The device structure was defined by subsequent photolithography and Ar ion-milling processes. Ohmic-contact electrodes and top gate electrode were Ti/Au deposited with an e-beam evaporator. Here, ion-milling was performed under 45 degree tilt condition on a rotating stage, resulting in the ramped side edge as schematically shown in Fig. 1a. This ensured the electrical contact to the top and bottom of the film.

**Acknowledgements**

R. Y. is supported by the Japan Society for the Promotion of Science (JSPS) through a research fellowship for young scientists. This research was supported by the Japan Society for the Promotion of Science through the Funding Program for World-Leading Innovative R & D on Science and Technology (FIRST Program) on "Quantum Science on Strong Correlation" initiated by the Council for Science and Technology Policy and by JSPS Grant-in-Aid for Scientific Research(S) No.24224009 and 24226002. This work was carried out by joint research of the Cryogenic Research Center, the University of Tokyo.

**Author contributions**

R. Y. performed thin films growth and device fabrication. R.Y., Y. K. and J. F. performed the low temperature transport measurements. R. Y. analyzed the data and wrote the manuscript with contributions from all authors. A. T., K. S. T., J. G. C., M. K. and Y. T.

jointly discussed the results and guided the project. Y. T. conceived and coordinated the project.

**Competing Financial Interests**

The authors declare no competing financial interests.

**Figure Legends**

**Figure 1 | Gating of topological insulator $(Bi_{0.16}Sb_{0.84})_2Te_3$ thin film. a, b,** Cross-sectional schematic and top-view photograph of a Hall-bar device. Broken line in (**b**) indicates the position for (**a**). **c,** Top gate voltage $V_G$ dependence of longitudinal conductance $\sigma_{xx}$ at various temperatures. **d, e,** Effective gate voltage ($V_G - V_{CNP}$) dependence of longitudinal and transverse resistance ($R_{xx}$ and $R_{yx}$) and inverse of Hall coefficient $1/R_H$ under magnetic field $B = 3$ T at temperature $T = 40$ mK. The $V_G$ for the charge neutral point (CNP), $V_{CNP}$, is defined at the gate voltage where $R_{yx}$ is crossing zero.

**Figure 2 | Observation of the quantum Hall effect. a, b, e and f,** Effective gate voltage $V_G - V_{CNP}$ dependence of $R_{xx}$ and $R_{yx}$ at various temperatures of $T = 40$-$700$ mK with application of magnetic field $B = 14$ T for the $x = 0.84$ (**a, b**) and $x = 0.88$ (**e, f**) films of

$(Bi_{1-x}Sb_x)_2Te_3$. **c, d, e and h,** Effective gate voltage $V_G - V_{CNP}$ dependence of $\sigma_{xx}$ and $\sigma_{xy}$ at various temperatures of $T = 40 - 700$ mK with application of magnetic field $B = 14$ T for the $x = 0.84$ and $x = 0.88$ films of $(Bi_{1-x}Sb_x)_2Te_3$ as deduced from the corresponding $R_{xx}$ and $R_{yx}$ data. Triangles in (**d**) and (**h**) show the dips of $\sigma_{xx}$. **i,** Schematics of the Landau levels (LLs) of the surface state of $(Bi_{1-x}Sb_x)_2Te_3$ ($x = 0.8$-$0.9$) thin film in case of the degenerate top and bottom surface states. At a high field, *e.g.* 14 T, the $n = -1$ LL of the surface state locates below the top of the valence band. When the Fermi energy ($E_F$) is tuned between the LLs, the quantum Hall state with index $\nu$ emerges. **j,** Schematics of the LLs of the top and bottom surface states in case the $n = 0$ (Dirac point) energy is different between the two surfaces.

**Figure 3 | Magnetic field dependence of Hall plateaus. a, b**, $\sigma_{xy}$ at $T = 40$ mK under various $B$ as a function of effective gate voltage $V_G - V_{CNP}$ for the $x = 0.84$ (**a**) and $x = 0.88$ (**b**) films. Traces for lower magnetic fields are each vertically offset by $e^2/h$. Dotted lines represent the plateau transitions as defined at $\sigma_{xy} = \pm 0.5\, e^2/h$. **c, d**, Quantized $\sigma_{xy}$ phase diagram for the $\nu = +1$, $0$ and $-1$ (blue, green and red shaded, respectively) states associated with the plateau edges (filled squares) and the plateau transition point (filled circles) defined at $\sigma_{xy} = \pm 0.5\, e^2/h$ in the plane of magnetic field and $V_G - V_{CNP}$. The

plateau edges are determined by the second-order $V_G$ derivative of $\sigma_{xy}$ (see SI).

**Figure 4 | Flows of $\sigma_{xy}$ and $\sigma_{xx}$.** ($\sigma_{xy}$, $\sigma_{xx}$) are displayed with the two experimental subparameters ($T$ and $V_G$). Each line connecting between points represents the flow behavior of ($\sigma_{xy}$, $\sigma_{xx}$) with lowering temperature from 700 mK to 40 mK at the specific value of $V_G$ at $B = 14$ T. The flows direct from upper to lower with decreasing temperature.

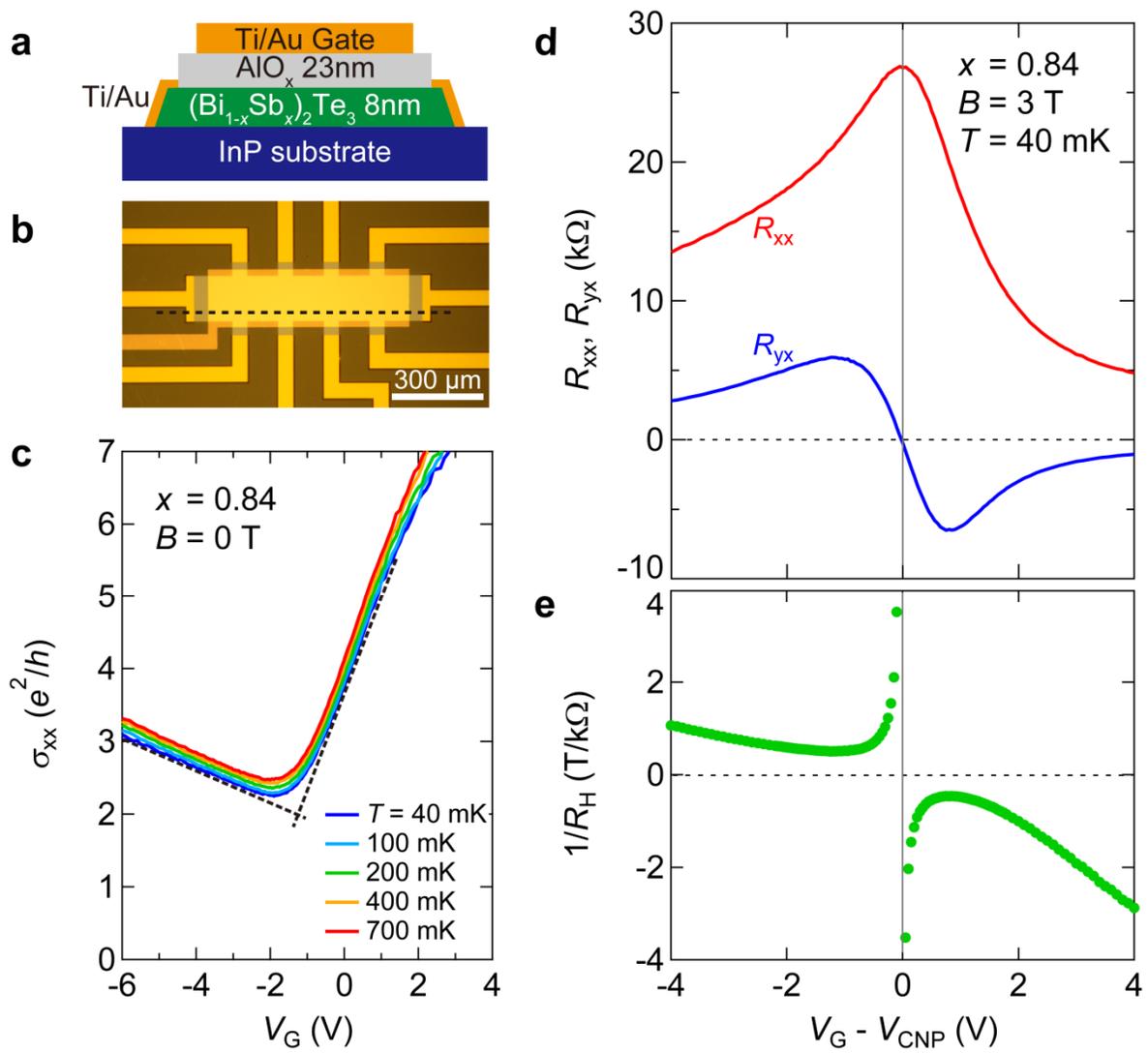

Fig. 1 R. Yoshimi *et al.*,

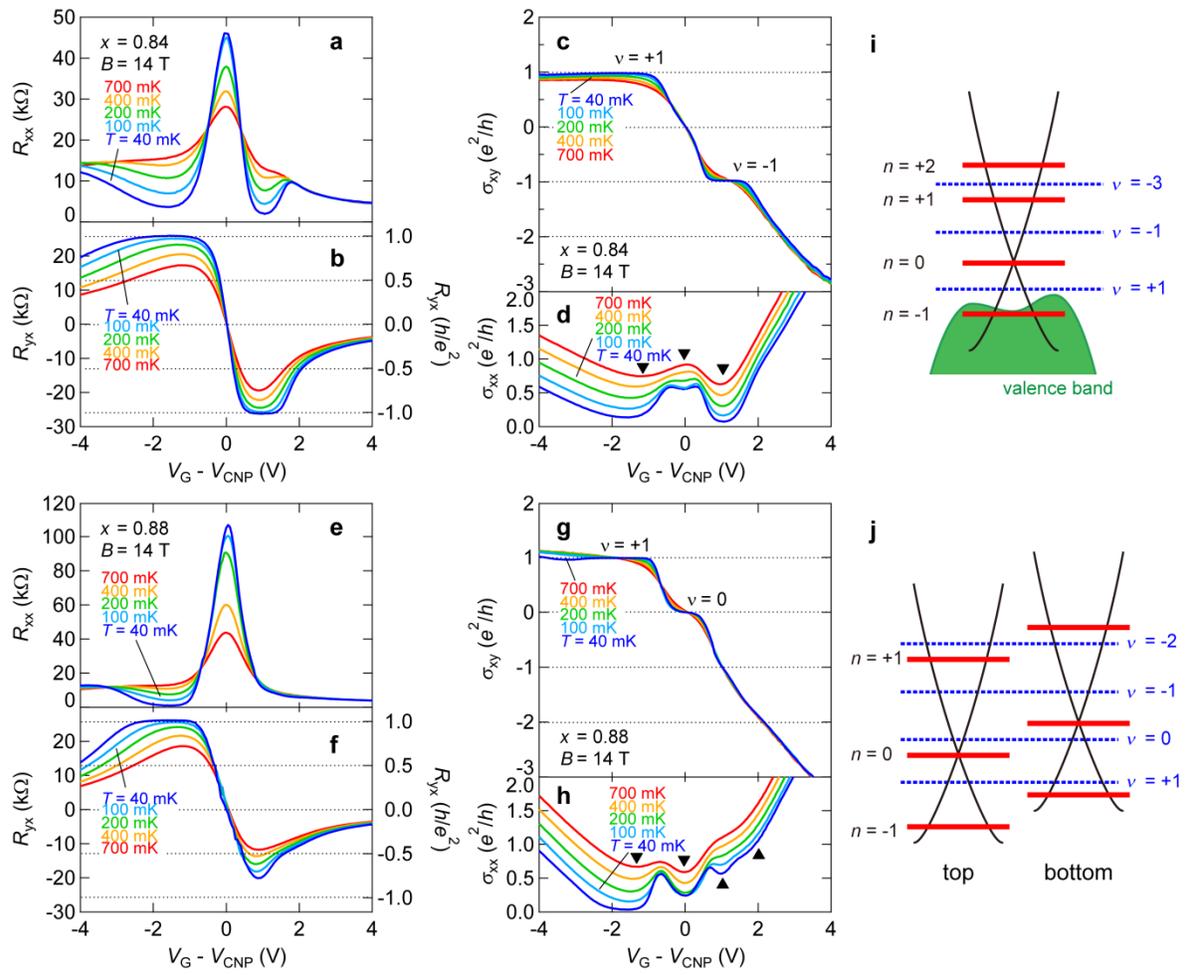

Fig. 2 R. Yoshimi *et al.*,

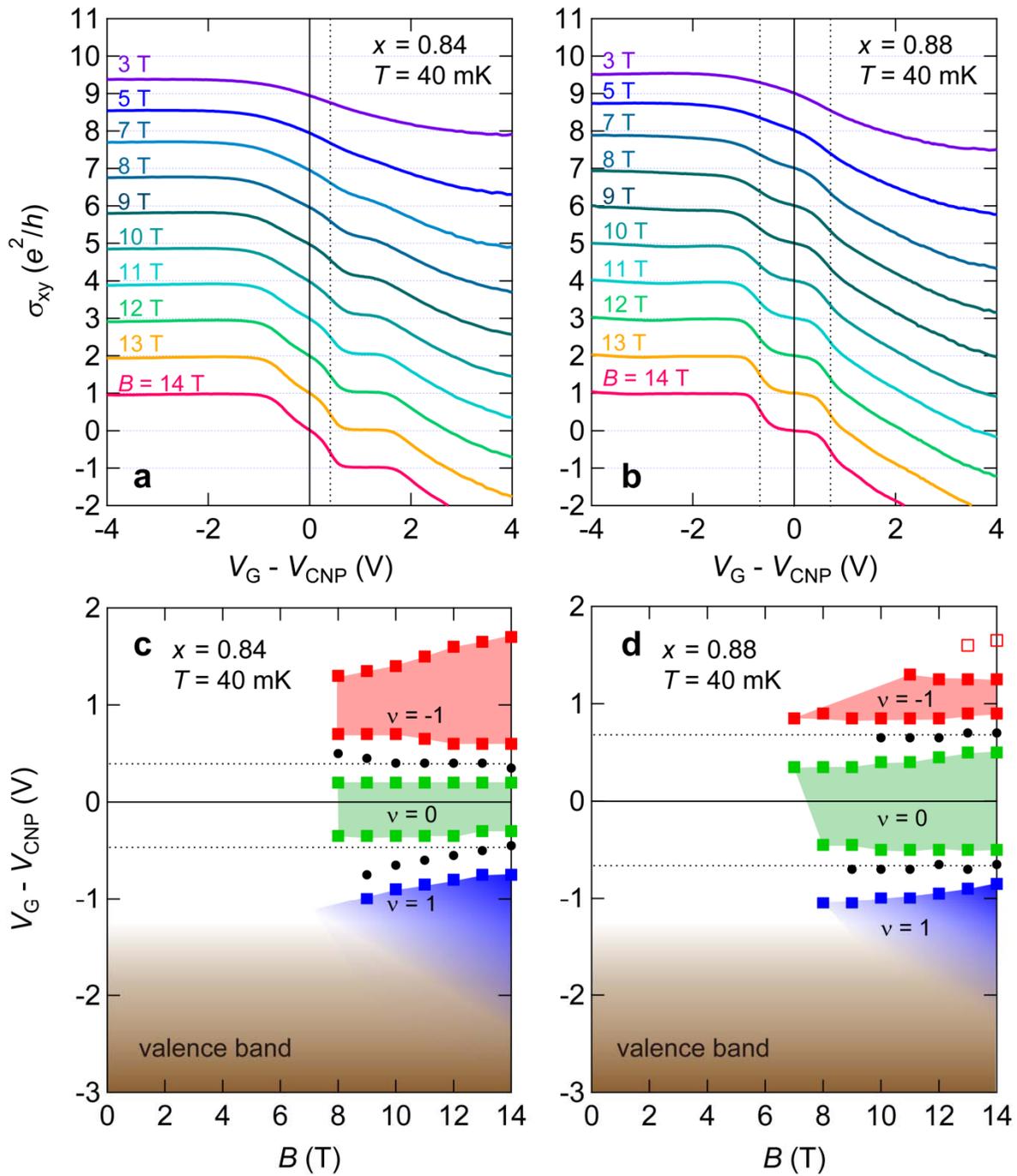

Fig. 3 R. Yoshimi *et al.*,

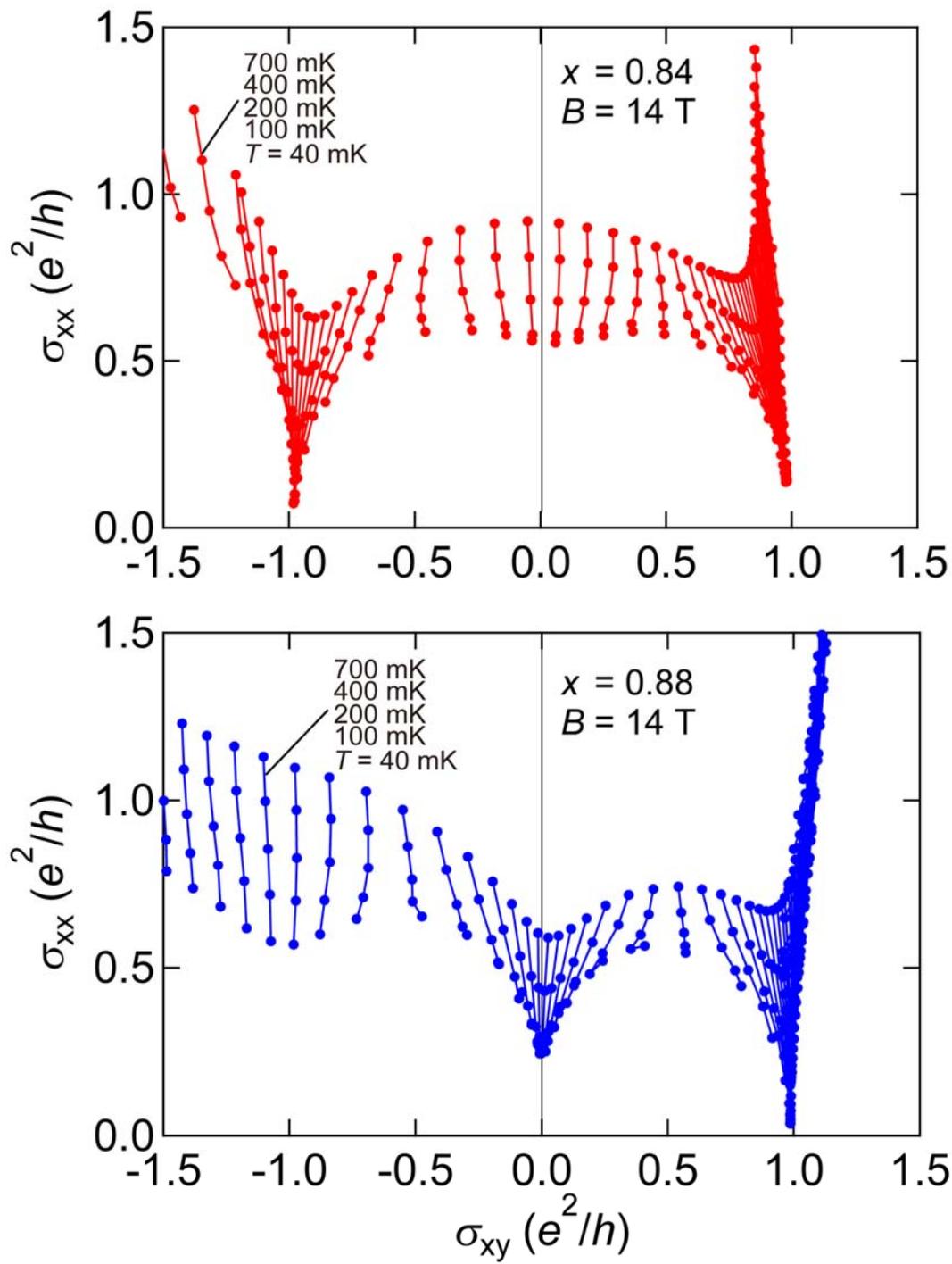

Fig. 4 R. Yoshimi *et al.*,

Supplementary Information for

**Quantum Hall Effect on Top and Bottom Surface States of Topological Insulator**

**$(Bi_{1-x}Sb_x)_2Te_3$ Films**

R. Yoshimi, A. Tsukazaki, Y. Kozuka, J. Falson, K. S. Takahashi, J. G. Checkelsky,

N. Nagaosa, M. Kawasaki and Y. Tokura

**1. Measurement configuration**

Low temperature transport measurements were performed in a dilution refrigerator with a superconducting magnet employing a lock-in technique at a low frequency (~ 3 Hz) and with a low excitation current (~ 1 nA). In the measurement circuit shown in Fig. S1. A series resistance of 1 GΩ was introduced to maintain a constant current condition that was confirmed by the signal from the lock-in amplifier that measured the current fed through the sample. We stress that all of the displayed data in the Figures in main text are the raw data under each condition without any (anti)symmetrization procedure. We did observe negligible asymmetry under positive and negative magnetic fields.

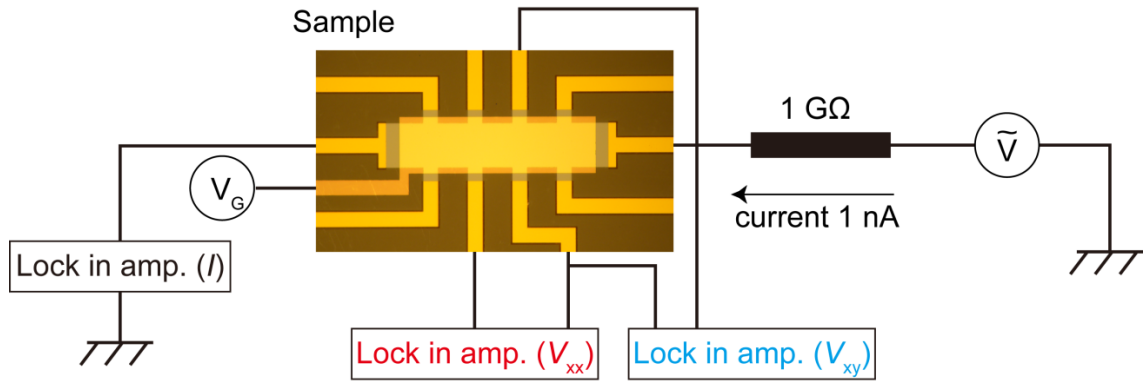

Fig. S1 Measurement configurations with an AC excitation.

## 2. Charge neutral point in the device ($x = 0.88$)

Transport characteristics for the device ($x = 0.88$) at $B = 0$ and 3 T are shown in Fig. S2. $V_G$ dependence of $\sigma_{xx}$ at $B = 0$ T presents similar trend to that in the device ($x = 0.84$) shown in Fig. 1c, exhibiting weak temperature dependence. It is noted that the minimum conductance was observed at around $V_G = 1$ V, indicating that the device ($x = 0.88$) presented here has originally *p*-type charge carriers, while the device ($x = 0.84$) presented in main text has *n*-type conduction. With applying $B = 3$ T, we obtained ambipolar behavior with a peak in $R_{xx}$ and a zero-crossing in $R_{yx}$. Contrary to the case of $x = 0.84$ device which has no $Sb_2Te_3$ buffer layer (see Fig. 1d in the main text), these two bias voltages differ slightly. Since we observe this difference only when a monolayer $Sb_2Te_3$ buffer is employed, we conclude that the difference in bias voltages is originating from the energy difference of the top and bottom surface states as depicted in the insets of Figs. S2b and S2c. At the bias voltage of $R_{yx} = 0$, electrons and holes exist on the top and bottom surface, respectively. Since the total conductivity is the sum of those at top and bottom surfaces, bias voltage at $R_{xx}$ maximum (conductivity minimum) does not correspond to that at $R_{yx} = 0$ owing to the mobility difference between electrons and holes. The energy difference between the two Dirac points at the top and bottom surface states is estimated to be $\delta E_{DP} \sim 70$ meV (see next section). We define $V_{CNP}$ for the device ($x = 0.88$) as the $V_G$ at the zero-crossing point in $R_{yx}$.

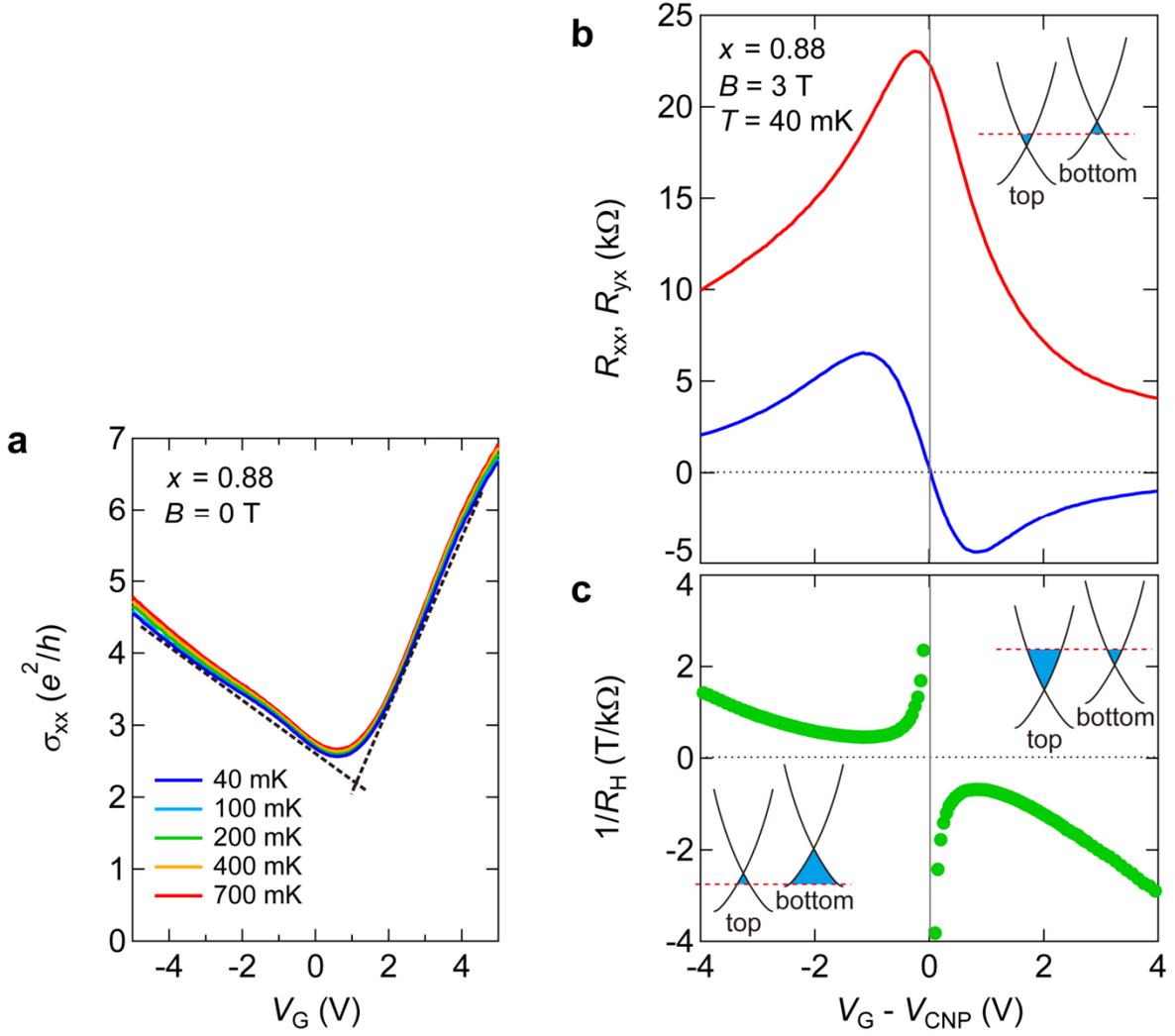

Fig. S2 **(a)** Top gate voltage ($V_G$) dependence of longitudinal conductance $\sigma_{xx}$ at various temperatures for the $x = 0.88$ device. The crossing bias position of two linear broken lines may be considered as the charge neutral point (CNP). However, the $V_G$ is slightly shifted from the $V_G$ giving CNP defined by Hall measurements. **(b), (c)** Effective gate voltage ($V_G$-$V_{CNP}$) dependence of longitudinal and transverse resistance ($R_{xx}$ and $R_{yx}$) and inverse of Hall coefficient $1/R_H$ under a magnetic field $B = 3$ T at a temperature $T = 40$ mK. $V_{CNP}$ is defined at the $V_G$ where $R_{yx}$ changes its sign. The inset schematics in (b) illustrate the Fermi level position between the Dirac points of the top and bottom surface states, representing the CNP. Right (left) inset in (c) indicates the Fermi level position that allows electron (hole) occupation on both surface states.

## 3. Energy scales in the band alignment.

To depict the relationship of Dirac dispersions on the top and bottom surface states shown in Fig. 2j, we deduce the energy difference of Dirac points ($\delta E_{DP}$) from the difference in the gate voltages ($\delta V_G$) of 1.4 V giving the plateau transitions from $\nu = 0$ to neighboring $\nu = \pm 1$ states indicated by dotted lines and black circles in Fig. 3d of the main text for the $x = 0.88$ device. In order to convert $\delta V_G$ to the change in charge carrier density $\delta n_{2D}$, we assumed $\delta n_{2D}\, e = C_{eff}\, \delta V_G$, where $C_{eff}$ is the effective capacitance. $C_{eff}$ is deduced as $5 \times 10^{11}$ cm$^{-2}$V$^{-1}$ from the linear part of $1/R_H$ vs. ($V_G - V_{CNP}$) relation shown in Fig. S2c. Then $\delta n_{2D}$ is estimated as $7 \times 10^{11}$ cm$^{-2}$. We adopted here the electron accumulation side for fitting, since overlap of valence band may give error in the estimation in hole accumulation side.

Assuming Dirac $k$-linear dispersion, *i.e.*, $E = \hbar v_F k$ and $n = k^2/4\pi$ (where $n$ is the carrier density on the single Dirac cone, $v_F$ is Fermi velocity and $k$ is Fermi wave number), $\delta n_{2D}$ can be converted to $\delta E_{DP}$. We took $v_F \sim 5 \times 10^5$ m/s as an averaged value of those of the electron and the hole in (Bi$_{0.12}$Sb$_{0.88}$)$_2$Te$_3$ (*1,2*) and obtained $\delta E_{DP} \sim 70$ meV for the $x = 0.88$ device. The similar procedure was adopted for the estimate of $\delta E_{DP}$ of the $x = 0.84$ device; from the value of $\delta V_G = 0.9$V (see Fig. 3c), we obtain $\delta E_{DP} \sim 50$ meV. The $\delta V_G$ value is in good agreement with the difference in $V_G$ giving maxima in $|R_{yx}|$ shown in Fig. S2c ($x = 0.88$) and Fig. 1e ($x = 0.84$). In between these maxima, electrons and holes coexist in the two Dirac states and the contribution to $R_{yx}$ cancels each other. When the Fermi level moves across the Dirac points, only electrons (holes) contribute to $R_{yx}$, giving maximum in $|R_{yx}|$.

## 4. Determination of the plateau edge in Fig. 3c and 3d.

The plateau edges represented by the solid squares in Figs. 3c and 3d in main text were defined from the second derivative of $\sigma_{xy}$ with respect to the $V_G$ as shown in Fig. S3. Under the magnetic field larger than 7 T, peaks and dips (lower triangles) appear as indicated, which correspond to the inflection points of $\sigma_{xy}$. Below 3T, no significant feature is observed.

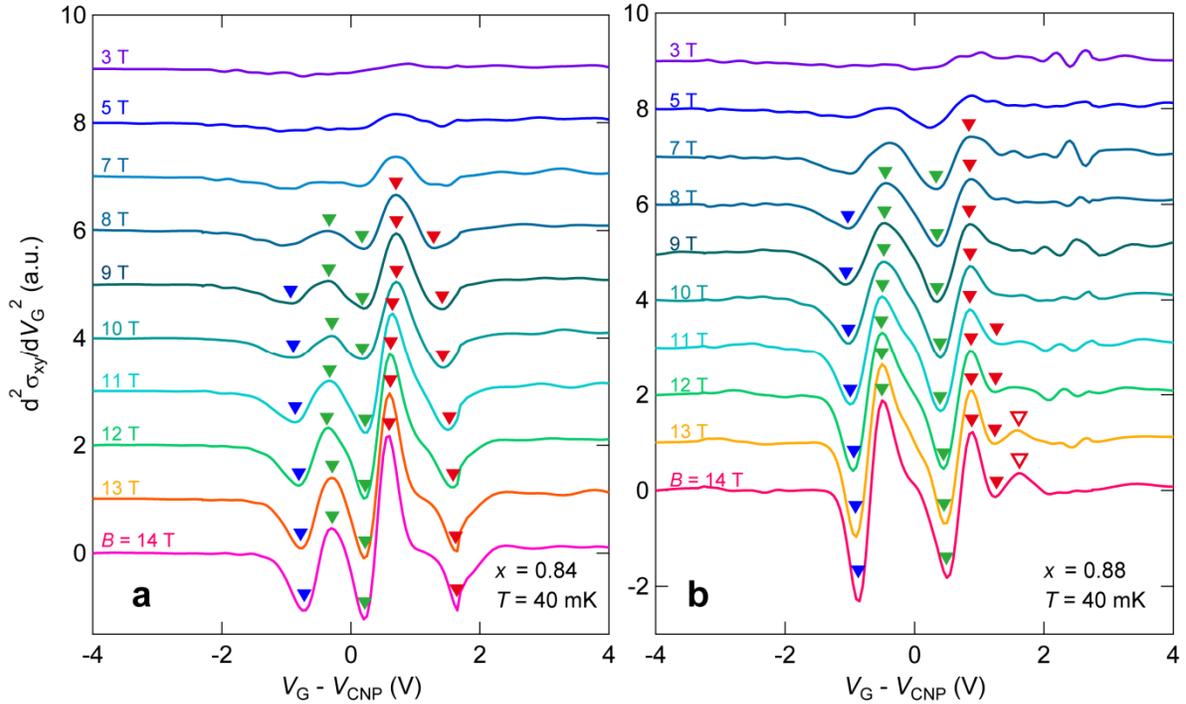

Fig. S3 Second derivative of $\sigma_{xy}$ with respect to the effective gate voltage $V_G - V_{CNP}$ at various magnetic fields for **(a)** $x = 0.84$ and **(b)** $x = 0.88$ devices. The colors of triangles correspond to those of the data plot in Fig. 3c and 3d.

## 5. Magnetic field dependence of $R_{xx}$ and $R_{yx}$ towards the QH state in the device

Figures S4 shows the magnetic field dependence of $R_{xx}$ and $R_{yx}$ at $T = 40$ mK. $V_G - V_{CNP}$ was tuned at the position to observe the largest $R_{yx}$ values. With increasing magnetic field, $R_{yx}$ linearly increases and bends to saturate at the quantum resistance $h/e^2$ at high magnetic field for $\nu = \pm 1$ of the $x = 0.84$ device (a) and (b) as well as for $\nu = +1$ of the $x = 0.88$ device (c). In this experiment, we have never observed Shubnikov-de Haas oscillation or quantum Hall plateaus at filling factor $\nu > 1$. This is attributed to low mobility of our films and positive magnetroresistance appearing under low magnetic field. The Quantum Hall plateau with $\nu = 2$ and 3 would have appeared at $B = 4$ and 5 T. At such a low magnetic field, a sharp increase in $R_{xx}$ is observed mainly due to the weak antilocalization which is common behavior in this system (3). As a result, $\omega_C \tau$ ($\omega_C$ is the cyclotron frequency and $\tau$ is the scattering time) is lower than 1, meaning appearance of quantum transport is not expected. Above 6 T, $R_{xx}$ starts to decrease and approaches zero. In case of the electron accumulation side in the $x = 0.88$ device, $R_{yx}$ does not reach to $h/e^2$ at $\nu = -1$ as shown in Fig.S4d and $R_{xx}$ does not display a minimum within the maximum available field ($B = 14$T).

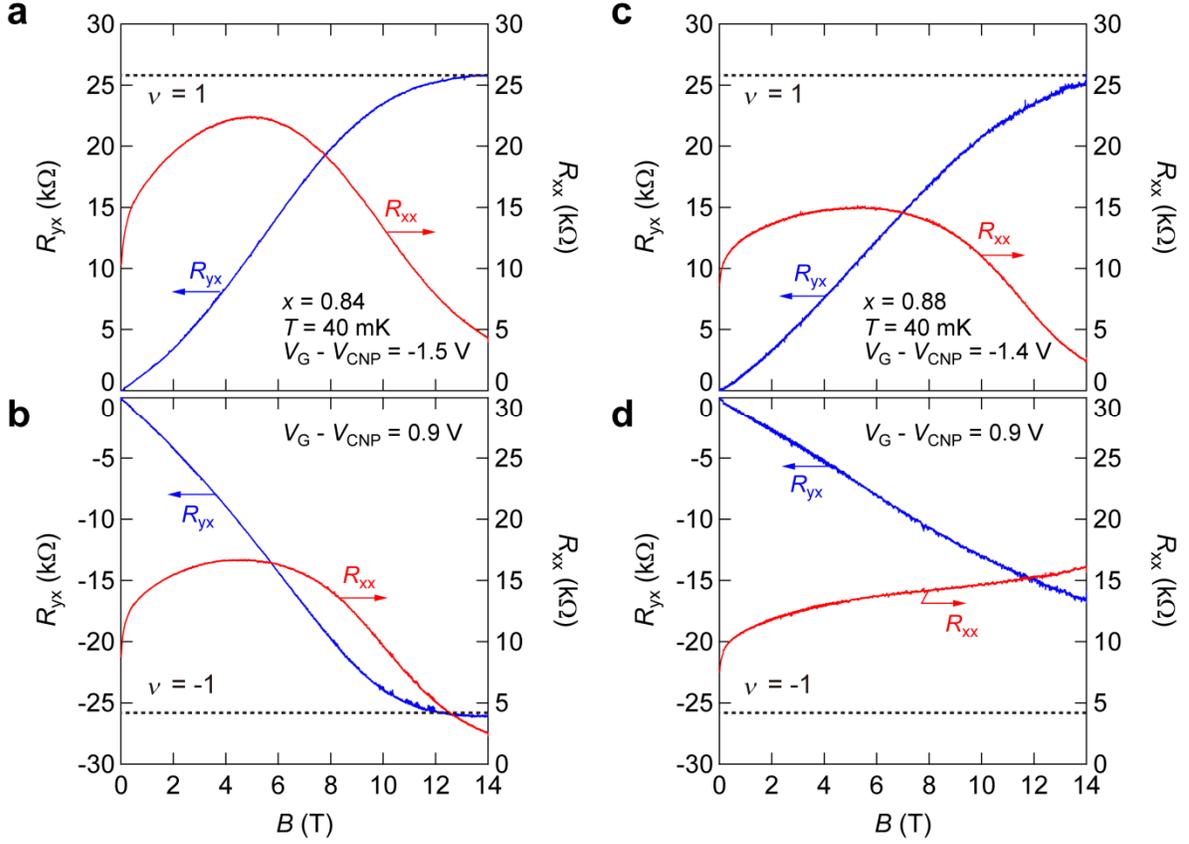

Fig.S4 Magnetic field dependence of $R_{xx}$ (red) and $R_{yx}$ (blue) at $T = 40$ mK for $x = 0.84$ (**(a)** and **(b)**) and for $x = 0.88$ (**(c)** and **(d)**) devices. $V_G$ is tuned at each QH state. Dashed lines indicate $\pm h/e^2 \sim 25.8$ k$\Omega$.